\documentclass[a4paper,12pt]{article}
\usepackage[utf8]{inputenc}
\usepackage[T1]{fontenc}
\usepackage{subfig}
\usepackage{hyperref}
\usepackage{color,soul}
\usepackage[normalem]{ulem}
\usepackage{amsmath,amssymb,amsfonts}
\usepackage{cite}
\usepackage[compress,numbers]{natbib}
\usepackage[pdftex,dvipsnames]{xcolor}
\usepackage{graphicx}
\usepackage{caption}
\usepackage{titlesec}
\usepackage{mathpazo}
\usepackage{txfonts}

\numberwithin{equation}{section}
\titleformat{\section}{\normalfont\sffamily\Large\bfseries}{\thesection}{1em}{}
\titleformat{\subsection}{\normalfont\sffamily\large\bfseries}{\thesubsection}{1em}{}

\DeclareMathAlphabet{\mathpzc}{OT1}{pzc}{m}{it}

\begin{document}
\begin{center}
{\Large{{\bf{\textsf{Setting the pace of microswimmers: when increasing viscosity speeds up self-propulsion}}}}}
\end{center}
\begin{center}
\vskip6pt
Jayant Pande$^{1,2}$, Laura Merchant$^{1,2,3}$, Timm Kr{\" u}ger$^{4}$, Jens Harting$^{5,6}$ and Ana-Sun{\v c}ana Smith$^{1,2,7\dagger}$

\vskip12pt

$^1${\footnotesize{\emph{Cluster of Excellence: EAM, Friedrich-Alexander University Erlangen-N\"urnberg, N\"agelsbachstra\ss e 49b, 91054 Erlangen, Germany}}}\\
$^2${\footnotesize{\emph{PULS Group, Department of Physics, Friedrich-Alexander University Erlangen-N\"urnberg, N\"agelsbachstra\ss e 49b, 91054 Erlangen, Germany}}}\\
$^3${\footnotesize{\emph{School of Physics and Astronomy, University of St. Andrews, North Haugh, St. Andrews KY16 9SS, The United Kingdom}}}\\
$^4${\footnotesize{\emph{School of Engineering, The University of Edinburgh, Edinburgh EH9 3JL, The United Kingdom}}}\\
$^5${\footnotesize{\emph{Helmholtz Institute Erlangen-Nuremberg for Renewable Energy (IEK-11), Forschungszentrum Jülich, F{\" u}rther Straße 248, 90429 Nuremberg, Germany}}}\\
$^6${\footnotesize{\emph{Department of Applied Physics, Eindhoven University of Technology, P.O.\ Box 513, 5600MB Eindhoven, The Netherlands}}}\\
$^7${\footnotesize{\emph{Division of Physical Chemistry, Ruđer Bošković Institute, Bijeni\v{c}ka cesta 54, Zagreb, Croatia}}}\\
\vskip4pt
$^\dagger${\footnotesize{Author for correspondence. E--mail:\ \href{mailto:smith@physik.fau.de}{smith@physik.fau.de}. }}
\vskip18pt
\end{center}

\begin{abstract}
It has long been known that some microswimmers seem to swim counter-intuitively faster when the viscosity of the surrounding fluid is increased, whereas others slow down. This conflicting dependence of the swimming velocity on the viscosity is poorly understood theoretically. Here we explain that any mechanical microswimmer with an elastic degree of freedom in a simple Newtonian fluid can exhibit both kinds of response to an increase in the fluid viscosity for different viscosity ranges, if the driving is weak. The velocity response is controlled by a single parameter $\varGamma$, the ratio of the relaxation time of the elastic component of the swimmer in the viscous fluid and the swimming stroke period. This defines two velocity-viscosity regimes, which we characterize using the bead-spring microswimmer model and analyzing the different forces acting on the parts of this swimmer. The analytical calculations are supported by lattice-Boltzmann simulations, which accurately reproduce the two velocity regimes for the predicted values of $\varGamma$.
\end{abstract}




It was discovered a few decades ago that many micro-organisms swim faster in more viscous fluids than in less viscous ones. In the first such finding, Shoesmith \cite{Shoesmith:1960:JGenM} reported the increased motility of \emph{Pseudomonas viscosa}, \emph{Bacillus brevis} and \emph{Escherichia coli} for a small increase in the viscosity of the solution; larger increases led to the motility decreasing. Similarly, Schneider and Doetsch \cite{Schneider:1974:JBact} reported that many flagellated bacteria showed an increase in the velocity when the solution viscosity rose to a characteristic value, and a decrease thereafter. Many other studies \cite{Kaiser:1975:Nature, Klitorinos:1993:Oral, Ruby:1998:FML, Nakamura:2006:BiophysJ} have corroborated this remarkable phenomenon, which gainsays both the intuitive expectation of a more viscous fluid providing greater resistance to motion, and the traditional theories of microbial motion in simple fluids all of which predict the velocity to go down with the viscosity \cite{Chwang:1971:RSB, Azuma:1992:Springer, Ramia:1993:BiophysJ}.   

Theoretical explanations in the past have focused on the non-Newtonian nature of the fluid and the structure of any polymers present therein, such as the possibility of the latter forming networks inside the fluid which facilitate swimmer propulsion \cite{Berg:1979:Nature, Magariyama:2002:BiophysJ, Nakamura:2006:BiophysJ, Leshansky:2009:PRE}. These mechanisms certainly contribute to the anomalous increase of swimmer velocity with fluid viscosity, yet they only concern particular combinations of microswimmer and fluid without attempting to explain the phenomenon in general. Moreover, such explanations suggest that the complex nature of the fluid is essential for the phenomenon to occur. 

Here we propose the opposite, by generalising the explanation to simple, structureless, Newtonian fluids. The central need for complexity in the fluids in the aforementioned explanations lies in the importance of having an interplay between two different time scales in the problem, one stemming from the elastic relaxation within the fluid and the other from the swimming cycle period (defined by the swimming stroke, which is the sequence of shapes that the swimmer adopts in order to propel its motion). Having a fast swimming stroke is not productive if the fluid itself does not relax before the succeeding swimming cycle can commence, and this leads to the existence of an optimal fluid viscosity for a swimmer with an assumed fixed swimming stroke rate. The same reasoning, however, should fit equally well the motion of a swimmer within a simple Newtonian fluid, \emph{as long as there is elasticity in the swimmer body itself}. Then the relaxation of the complex fluid can be replaced by the relaxation of any body deformations within the swimmer in the viscous fluid, which again interacts with the stroke time scale to lead to different velocity responses to an increase in the fluid viscosity. For mechanically driven microswimmers, an \emph{effective elasticity} can be defined assuming the body deformations (including the beating of appendages such as flagella) occur at a steady rate, meaning that they should exhibit both kinds of velocity vs.\ viscosity response.


This argument does not apply to microswimmers whose swimming stroke is predefined, independent of the fluid's influence, as is often the case for theoretical microswimmer models \cite{Najafi:2004:PRE, Purcell:1977:AmJPhys, Taylor:1951:PRSLA, Avron:2005:NewJPhys, Lighthill:1952:CPAM, Ledesma-Aguilar:2012:EPJE, Lauga:2008:PRE}. It is well-known that the distance covered by a microswimmer in one swimming cycle is proportional to the area of the closed loop in configuration space that its swimming stroke describes \cite{Shapere:1987:PRL, Shapere:1989:JFM}. The effect of the different forces acting on the swimmer is subsumed in the swimming stroke, meaning that when the latter is imposed then the effect of force parameters such as the viscosity is lost. (As illustration, see the velocity expressions for three prominent microswimmer models in \cite{Taylor:1951:PRSLA, Golestanian:2008:PRE, Lighthill:1952:CPAM}, each of which depends solely on the respective swimmer's geometrical parameters.) Hence, to see the full dependence of the swimming velocity on the fluid viscosity, a force/energy-centric approach, which allows the swimmer to adjust its swimming stroke in response to the driving, is necessary.

To test the above general argument
, we perform an analytical and numerical study of a mechanical microswimmer in a simple Newtonian fluid, based on the popular three-sphere model of Najafi and Golestanian \cite{Najafi:2004:PRE}. In the three-sphere model the swimming stroke is imposed and the swimmer consequently does not exhibit a velocity dependence on the fluid viscosity \cite{Najafi:2004:PRE, Golestanian:2008:PRE}. Our purposes therefore require us to modify this model, by including springs between the spheres and imposing the forces driving the motion instead of the swimming stroke, allowing the latter to emerge in response to the former. 
This reworked model 
is amenable to fully analytical treatment, yet is simple enough--with the motion being driven only by two elastic degrees of freedom--to allow one to generalize the results to other microswimmers which are driven by elastic components.

Analysis of the model confirms the fact that two regimes of motion exist, in one of which the swimmer gets slower (which we call the `conventional' regime) and in the other one faster (the `aberrant' regime) when the viscosity of the surrounding fluid is increased. The regimes depend on a ratio $\varGamma$ of two characteristic time scales,
\begin{equation}
\varGamma =  \dfrac{\text{relaxation time of spheres in fluid}}{\text{swimming cycle period}}. 
\end{equation}
Assume that the swimming cycle period is fixed. For $\varGamma \gg 1$ the spheres do not relax fully within one swimming cycle, and increasing the fluid viscosity causes them to relax even less, making the swimmer swim slower. For $\varGamma \ll 1$ the spheres relax very quickly, but that is not advantageous since the swimming speed is limited by the cycle period. Moreover, a quick relaxation rate of the spheres (concomitant with a small fluid viscosity) also reduces the inter-sphere hydrodynamic interactions which are vital to the swimming motion. Therefore in this case increasing the fluid viscosity leads to faster swimming. This whole picture can be seen from the reverse point of view, where $\varGamma$ is modified by changing the swimming cycle period instead of the sphere relaxation time. As shown in a previous work \cite{Pande:2015:SoftM}, the swimming velocity shows a maximum as a function of the cycle frequency, because for very quick driving the spheres do not relax within one cycle, and in the limit of very slow driving the swimming tends to cease. Note that this rate-dependence of the swimming velocity disappears in the Stokesian realm once the stroke is imposed (apart from an overall scaling factor of the frequency), since there is then only one characteristic time scale in the problem.


The dependence of the aberrant swimming phenomenon on the sphere (or, more generally, body deformation) relaxation time speaks immediately to the necessity of having an elastic degree of freedom in the swimmer which couples to the fluid viscosity to determine the rate of relaxation. Lastly, since the aberrant regime is observed for small viscosities, then for the low Reynolds number condition of microswimming to be honored, the driving (and hence the motion velocity) needs to be weak. 

\subsection*{Analytical swimmer model}


\begin{figure}
\centering
\includegraphics[width=0.42\textwidth]{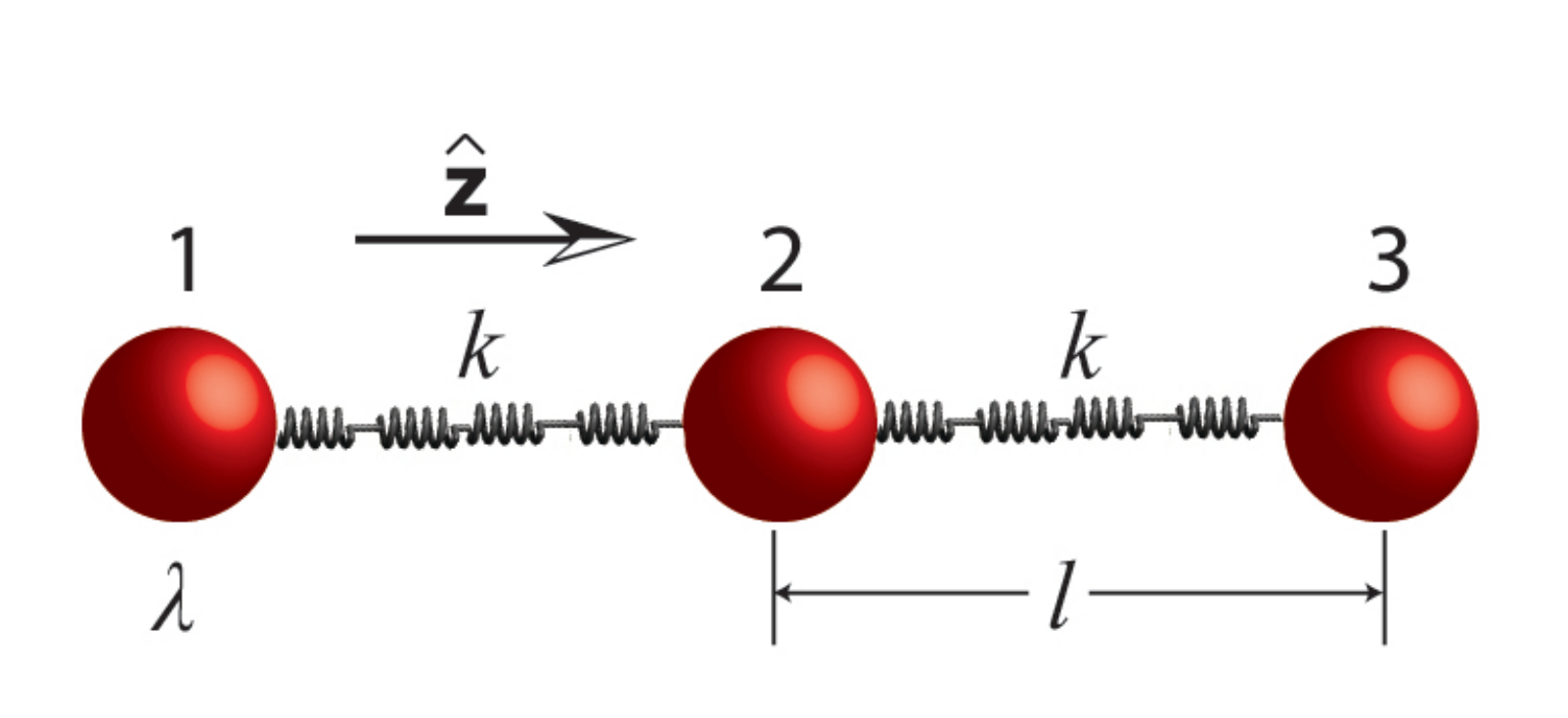}
\caption{(color online) Swimmer model with springs. $\lambda$ is the reduced friction coefficient of the beads (or the radius for spherical beads).}\label{fig:swimmer}
\end{figure}

\begin{figure*}
\centering
\includegraphics[width=0.9\textwidth]{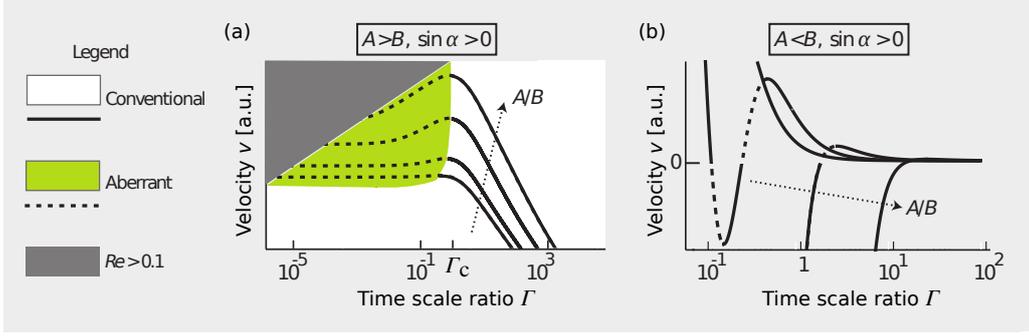}
\caption{(color online) Velocity vs.\ viscosity curves (with $\varGamma$ here acting as a dimensionless viscosity, from Eq.~(\ref{eq:Gamma})) with $\sin\alpha > 0$ and different force amplitude ratios ($A/B > 1$ in (a) and $A/B < 1$ in (b)). The values of the different parameters have been kept physically appropriate for microswimming. Note that in (b), where the swimmer velocity changes sign as a function of the viscosity, the conventional and the aberrant parts of the curves are defined with respect to the magnitude of the velocity.}\label{fig:v-eta}
\end{figure*}

The swimmer consists of three beads connected in series by two harmonic springs (Fig.~\ref{fig:swimmer}), and driven by known forces of the form
\begin{align}\label{eq:dr_forces}
&\mathbf F_1^\text{d}(t) = A \sin\left(\omega t\right) \mathbf{\hat{z}},\nonumber\\ 
&\mathbf F_2^\text{d}(t) = - \mathbf F_1^\text{d}(t) - \mathbf F_3^\text{d}(t),\text{ and}\nonumber\\
&\mathbf F_3^\text{d}(t) = B \sin\left(\omega t + \alpha\right) \mathbf{\hat{z}},\text{ with }\alpha \in[-\pi, \pi].
\end{align}
Here $A$ and $B$ are non-negative amplitudes of the time-dependent driving forces $\mathbf F_1^\text{d}(t)$ and $\mathbf F_3^\text{d}(t)$ applied along the $\mathbf{\hat{z}}$-direction
 to the outer beads at the frequency $\omega$ and with the phase difference $\alpha$. The force $\mathbf F_2^\text{d}(t)$ on the middle bead is set by the condition for autonomous propulsion, which requires the net driving force on the device to vanish at all times. The two springs are identical, with a stiffness constant $k$ and a rest length $l$ which is much larger than the bead dimensions. For convenience, we define a `reduced friction coefficient' $\lambda$ of the beads as 
\begin{equation}
\lambda = \dfrac{\gamma}{6 \pi \eta},
\end{equation}
where $\gamma$ is their Stokes drag coefficient and $\eta$ is the dynamic viscosity of the fluid. The parameter $\lambda$ has dimensions of length and plays the role of the radius for non-spherical beads \cite{Pande:2015:SoftM}.

For our swimmer the ratio of time scales $\varGamma$ becomes
\begin{equation}\label{eq:Gamma}
\varGamma = \tau_\text{s}\ \omega = \dfrac{6 \pi \omega \lambda \eta}{k},
\end{equation}
where $\tau_\text{s}$ is the relaxation time of the spheres in the fluid, defined as
\begin{equation}\label{eq:tau_s}
\tau_\text{s} = \dfrac{\gamma}{k} = \dfrac{6 \pi \lambda \eta}{k}.
\end{equation}

The fluid is assumed to be governed by the Stokes equation
\begin{equation}
\eta\nabla^2\mathbf u\left(\mathbf r, t\right) - \nabla p\left(\mathbf r, t\right) + \mathbf f\left(\mathbf r, t\right) = \mathbf 0,
\end{equation}
and the incompressibility condition 
\begin{equation}
\nabla\cdot\mathbf u=0.
\end{equation}
Here $\mathbf u\left(\mathbf r, t\right)$ and $p\left(\mathbf r, t\right)$ are the velocity and the pressure of the fluid at the point $\mathbf r$ at time $t$. The force density $\mathbf f\left(\mathbf r, t\right)$ acting on the fluid, in the limit of small bead dimensions, is given by
\begin{equation}
\mathbf f\left(\mathbf r, t\right)=\sum\limits_{i=1}^3\left[\mathbf F^\text{d}_i(t) + \mathbf F^\text{s}_i(t)\right]\delta\left(\mathbf r - \mathbf R_i(t)\right),
\end{equation}
where the index $i=1, 2, 3$ denotes the $i$-th bead placed at the position $\mathbf R_i(t)$ subject to a driving force $\mathbf F^\text{d}_i(t)$ and a spring force $\mathbf F^\text{s}_i(t)$ (which, for the middle bead, results from two springs). Here $\delta\left(\mathbf r\right)$ denotes the Dirac delta function. Assuming no slip at the fluid-bead interfaces, the instantaneous velocity $\mathbf v_i(t)$ of each bead \cite{Doi:1988:OUP} is given by
\begin{equation}\label{eq:v_bodies}
\mathbf v_i=\dfrac{\mathrm d\mathbf R_i}{\mathrm dt} = \left(\mathbf F^\text{d}_i + \mathbf F^\text{s}_i\right)\gamma^{-1} + \sum\limits_{j \neq i}\mathbf T\left(\mathbf R_i - \mathbf R_j\right)\cdot\left(\mathbf F^\text{d}_i + \mathbf F^\text{s}_i\right),
\end{equation}
where $\mathbf T\left(\mathbf r\right)$ is the Oseen tensor \cite{Happel:1965:P-H, Oseen:1927:LeipzigAV}, and is here diagonal due to the collinear nature of the driving forces and the employed far-field approximation (which assumes that the bead dimensions are much smaller than $l$).

In the steady state the bead positions are of the form \cite{Felderhof:2006:PhysFluids}
\begin{equation}\label{eq:Ri}
\mathbf R_i(t)=\mathbf S_{i0} +  \boldsymbol\xi_i(t) + \mathbf vt
\end{equation}
due to the sinusoidal nature of the forces. Here $\boldsymbol\xi_i(t)$ denotes small sinusoidal oscillations around the uniformly-moving equilibrium configuration $\mathbf S_{i0} + \mathbf vt$, where $\mathbf S_{i0}$ are the initial positions of the beads and $\mathbf v$ is the mean cycle-averaged uniform swimming velocity of the assembly. Clearly we have $|\mathbf S_{20} - \mathbf S_{10}| = |\mathbf S_{30} - \mathbf S_{20}| = l$.

\begin{figure*}
\centering
\includegraphics[width=0.94\textwidth]{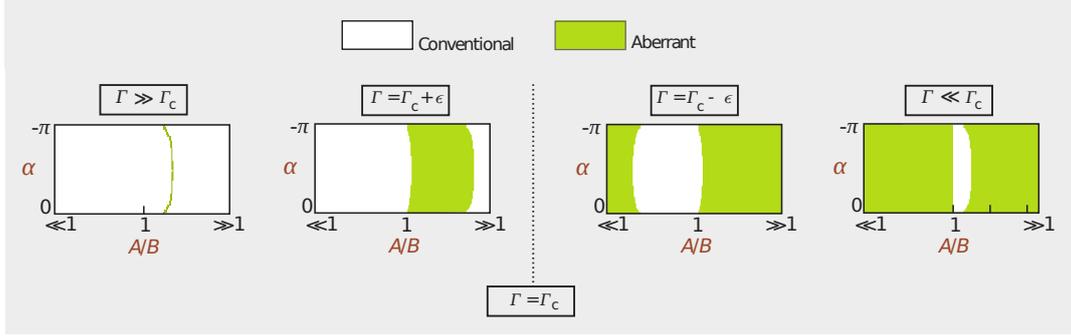}
\caption{(color online) Phase diagrams marking the conventional and the aberrant regimes of the swimmer's motion for different values of the forcing parameters $\alpha$, $A$ and $B$. The parameter $\varGamma$ is varied between the different plots by changing only the spring constant $k$.
}\label{fig:v-eta-phase}
\end{figure*}

Eqs.~(\ref{eq:v_bodies}) and (\ref{eq:Ri}) lead to a coupled system of differential equations in the $\boldsymbol\xi_i$'s, which can be solved by a perturbative scheme by expanding the arm-lengths $|\mathbf R_2(t) - \mathbf R_1(t)|$ and $|\mathbf R_3(t) - \mathbf R_2(t)|$ around their mean values $l$, if we assume that the driving forces, and consequently the oscillations $\boldsymbol\xi_i(t)$, are small, \emph{i.e.} $|\boldsymbol\xi_i(t)|\ll l$ for all $i$ and all times $t$ \cite{Felderhof:2006:PhysFluids}. Since the forces and the displacements are all sinusoidal, the first-order terms in the perturbation variable $\boldsymbol\xi_i(t)$ turn out to be zero. We calculate to the second order in $\boldsymbol\xi_i/l$, therefore, and find the velocity expression for the swimmer to be
\begin{align}\label{eq:v_force}
&\mathbf v = \nonumber\\
&\dfrac{7 \omega \lambda \left[A B\left(k^2 + 12\pi^2 \omega^2 \eta^2 \lambda^2 \right)\sin\alpha + 2\pi\left(A^2 - B^2\right)k \omega \eta \lambda \right]}{24 l^2 \left(k^2 + 4 \pi^2 \omega^2 \eta^2 \lambda^2\right)\left(k^2 + 36 \pi^2 \omega^2 \eta^2 \lambda^2 \right)} \mathbf{\hat{z}}.
\end{align}
This expression, being a non-monotonic function of the viscosity $\eta$, 
directly shows the existence of the conventional and aberrant velocity-viscosity regimes.

\subsection*{Characteristics of the swimming regimes}

We now study these regimes by changing only the viscosity $\eta$ and keeping all the other independent parameters in the problem fixed, including the driving forces. An alternative approach would be to vary the driving forces such that the efficiency of the compared swimmers is held constant. 
Fixing the driving forces is easier, and the results for constant efficiencies would be essentially the same since the efficiencies of fast swimmers are generally higher than those of slow ones \cite{Pande:2015:SoftM, Pickl:2012:JoCS}.

We find that when $(A-B)/\sin\alpha>0$, then the velocity $\mathbf v$ as a function of $\eta$ has exactly one extremum (see Fig.~\ref{fig:v-eta}a, where $\varGamma$ plays the role of a dimensionless viscosity since all the factors except $\eta$ in Eq.~(\ref{eq:Gamma}) are held constant). This extremum divides the conventional regime, obtained for large viscosities and shown in white in Fig.~\ref{fig:v-eta}a, and the aberrant regime, obtained for small viscosities and shown in green (light gray in grayscale print). The different curves correspond to increasing values of $A$, with $B$ constant. 
Some manipulation of the velocity expression shows that for each curve the swimmer lies in the conventional regime if 
\begin{equation}\label{eq:psi-eta}
\varGamma > \varGamma_\text{c} = \dfrac{3}{\sqrt{5 + 2\sqrt{13}}} \approx 0.86.
\end{equation}

The dark gray area marks the region where the swimmer Reynolds number $Re > 0.1$ (where $ Re = |\mathbf v| l \rho/\eta$, with $\rho$ denoting the fluid density), when we assume the condition of Stokes flow to be violated. For large enough values of the driving force amplitude $A$, the entire part of the velocity curve for which $Re < 0.1$ falls in the conventional regime.

If  $(A-B)/\sin\alpha<0$, then the velocity can have several local extrema (Fig.~\ref{fig:v-eta}b). Moreover, the swimmer can reverse direction if the fluid viscosity is changed. If the force parameters satisfy the condition
\begin{equation}
\dfrac{B}{A} > (1 + 6 \sin^2\alpha + 2\sin\alpha\sqrt{3 + 9\sin^2\alpha})^{1/2},
\end{equation}
then the swimmer becomes aberrant for an intermediate range of viscosities (dashed parts of curves in Fig.~\ref{fig:v-eta}b).

\begin{figure*}
\centering
\includegraphics[width=0.9\textwidth]{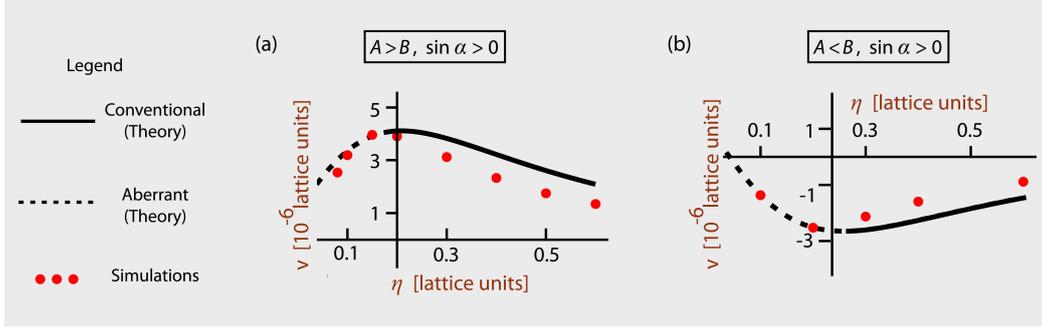}
\caption{(color online) Comparison of the swimming velocity as a function of the fluid viscosity from theory (solid and dashed curves) and simulations (red points), with $\sin\alpha > 0$ and (a) $A/B = 20$, and (b) $A/B = 0.04$. 
}\label{fig:v-eta_sim}
\end{figure*}

To depict the importance of the time scale ratio $\varGamma$ in controlling the regime of motion, we fix in Fig.~\ref{fig:v-eta-phase} the fluid viscosity $\eta$ and change $\varGamma$ by varying the spring stiffness $k$. The different plots in Fig.~\ref{fig:v-eta-phase} mark the conventional and the aberrant regimes for different values of the driving force parameters $A$, $B$ and $\alpha$, and for decreasing $\varGamma$. At large $\varGamma$ values, the conventional regime is dominant (leftmost panel in Fig.~\ref{fig:v-eta-phase}). In the limit of infinite $\varGamma$, which corresponds to zero elasticity ($k = 0$), the whole phase space is conventional (as is easily seen by putting $k = 0$ in Eq.~(\ref{eq:v_force})), since the spheres never relax back from any displacement and there is no competition of time scales in the problem. As $\varGamma$ decreases, the relative area of the aberrant regime rises continuously (center left panel in Fig.~\ref{fig:v-eta-phase})
as long as the inequality (\ref{eq:psi-eta}) is satisfied. At the critical value $\varGamma = \varGamma_c (\approx 0.86)$, there is a discontinuous change in the nature of the regimes across most of the phase space, with the aberrant regime becoming dominant (center right and right panels in Fig.~\ref{fig:v-eta-phase}), recalling our earlier discussion of the swimming being aberrant for small values of $\varGamma$. 

\subsection*{Lattice-Boltzmann simulations}

To confirm the existence of the two viscosity-dependent regimes that the theory predicts, we employ numerical simulations, which are a commonly-used tool to study microswimming (utilized, for instance, in \cite{Pooley:2007:PRL, Lauga:2009:RepProgPhys, Swan:2011:PhysFluids, Jansen:2011:PRE, Arroyo:2012:PNAS, Saintillan:2012:JRSI, Zoettl:2012:PRL, Elgeti:2013:PNAS, Kantsler:2013:PNAS, Marchetti:2013:RevModPhys, Lushi:2014:PNAS, Kaiser:2014:PRL, Qiu:2014:NatCommun, Zhan:2014:JFM, Zoettl:2014:PRL, Schaar:2015:PRL, Elgeti:2015:RepProgPhys}). For this we use the LB3D code \cite{Jansen:2011:PRE, Krueger:2013:EPJST} based on the immersed-boundary method (IBM) and the lattice-Boltzmann method (LBM) with a standard D3Q19 lattice and the BGK collision operator as described in \cite{Krueger:2011:CMA}. The beads are identical rigid spheres of radius $5 \Delta x$, where $\Delta x$ is the resolution of the lattice-Boltzmann fluid, and their surface is represented by $720$ immersed boundary points. The equilibrium center-to-center distance between the spheres is $l = 36 \Delta x$, and the spring constant equals $k = 0.02$ in lattice units. The simulations are run for $30$ cycles to let the undesired transients decay, with the period of each cycle being $8000$ time steps. The system size is $200\times80\times78\text{ }\Delta x^3$ and periodic boundaries are employed.

We run two sets of simulations, to account for both the cases of $(A - B)/\sin\alpha \gtrless 0$. For this we fix $\alpha = \pi/2$, and the ratio $A/B$ of the driving force amplitudes in the two sets is kept at $20$ and $0.04$. These driving forces on each bead are distributed evenly across all of its immersed surface points, and are always kept small enough so that the resulting Reynolds number is smaller than $0.1$ to ensure `low $Re$' swimming \cite{Pooley:2008:CPC}.


Figs.~\ref{fig:v-eta_sim}a and \ref{fig:v-eta_sim}b show the average swimmer velocity $\mathbf v$ in the steady state as a function of the tested viscosity values $\eta$, for the two force amplitude ratios. In both the investigated cases we observe that the two predicted velocity-viscosity regimes are reproduced well, 
with the small errors being attributable to the unrealisably small radius to arm-length ratios in the theoretical model, in addition to the limitations inherent in simulations (such as boundary effects and imperfect space and time discretization).

\subsection*{Conclusion}

With the help of a bead-spring swimmer model, we have explained on physical grounds the puzzling observation of some micro-organisms seeming to swim faster in more viscous fluids. We suggest that this is a more universal phenomenon than previously thought, with the velocity of any mechanical microswimmer rising and falling with the fluid viscosity in different viscosity ranges, as long as two conditions are satisfied: the swimming stroke is not too strong to preclude low Reynolds number swimming at the small viscosities where the aberrant regime is observed, and the swimmer possesses an elastic degree of freedom which may freely respond to external forces. Which regime the motion occurs in depends on the ratio of two characteristic time scales in the system, one determined by the relaxation of the swimmer deformations within the fluid and the other by the rate of applying these deformations. We have supported the analytical calculations with lattice-Boltzmann simulations, and shown that the simulations reproduce the velocity-viscosity regimes very well. Our work uncovers the fact that both these regimes are attained in simple Newtonian fluids at negligible Reynolds numbers, and by this provides fundamental insight into the way microswimming is affected by the interaction between the viscosity of the fluid and the elasticity of the swimmer.

\subsection*{Acknowledgements}
A.-S.~S. and J.~P.~thank the KONWIHR ParSwarm grant as well as the funding of the Deutsche Forschungsgemeinschaft (DFG) through the Cluster of Excellence Engineering of Advanced Materials. J.~H.~acknowledges support by NWO/STW (Vidi grant 10787) and L.~M.~thanks the DAAD for a RISE scholarship. T.~K.~thanks the University of Edinburgh for the award of a Chancellor's Fellowship. We are also very grateful to Fred MacKintosh for helpful comments and discussions. 


\bibliography{rsc3} 
\bibliographystyle{rsc3}

\end{document}